\documentclass[epj,nopacs]{svjour}
\usepackage{graphics}
\usepackage{epsfig}
\usepackage{amssymb}
\begin{document}
\title{Resolved ${\mathbf {\gamma^*_L}}$ in hard
collisions of virtual photons: QCD effects}
\author{Ji\v{r}\'{\i} Ch\'{y}la \and Marek Ta\v{s}evsk\'{y}
% etc
%\thanks is optional - remove next line if not needed
\thanks{Work done within the {\em Center for Particle Physics}
under the project LN00A006 of the Ministry of Education of the
Czech Republic.}                     % Do not remove
}
\institute{Institute of Physics
of the Academy of Sciences of the Czech Republic\\
Na Slovance 2, Prague 8, Czech Republic}
%\and the second here}
%
\date{Received: date / Revised version: date}
\abstract{
The manifestations of QCD effects on quark and gluon distribution
functions of longitudinally polarized virtual photons involved in hard
collisions are investigated. It is shown that for moderate photon
virtualities and in the kinematical region accessible at HERA and LEP
these effects are sizable and significantly enhance theoretical
predictions based on contributions of transversally polarized virtual
photon only.
%
%\PACS{
%      {PACS-key}{discribing text of that key}   \and
%      {PACS-key}{discribing text of that key}
%     } % end of PACS codes
} %end of abstract
\maketitle
\section{Introduction}
\label{intro}
In QED quantized in covariant gauge, longitudinally polarized
on--shell photons are present, but due to gauge invariance decouple,
order by order in perturbation theory, in expressions for physical
quantities. For the virtual photon with virtuality
\footnote{In this paper the virtuality of a particle with
four--momentum $k$ and mass $m$ is defined as $\mid k^2-m^2\mid$.}
$P^2$ its longitudinal polarization, denoted $\gamma_L^*$, does
contribute to physical quantities and gauge invariance merely
requires that these contributions vanish as $P^2\rightarrow 0$.
In a previous publication \cite{long} we have discussed the
contributions of $\gamma_L^*$ to two physical quantities using purely
QED formula for quark distribution functions of $\gamma_L^*$. In this
paper we continue our investigation of the relevance of $\gamma_L^*$ in
hard collisions by incorporating the effects of QCD radiation on parton
distribution functions (PDF) of $\gamma_L^*$ recently derived in
\cite{plb}. In the next Section the rationale for introducing the
concept of the structure of virtual photon is recalled, followed
in Sections 3 and 4 by a short review of the QED and QCD formulae for
corresponding PDF. The numerical relevance of the contributions of
resolved $\gamma_L^*$ with QCD improved PDF are discussed in Sections
5 for the LO and in Section 6 for the NLO QCD calculations.
\section{Virtual photon and its ``structure''}
\label{sec:structure}
Let us briefly recall the virtue of extending the concept of
partonic ``structure'' to virtual photons \cite{prd,friberg}:
\begin{itemize}
\item In principle, the concept of partonic structure of
virtual photons can be dispensed with as higher order QCD corrections
to cross sections of processes involving virtual photons in the initial
state are well--defined and finite even for massless partons.
\item In practice, however, the concept of {\em resolved virtual
photon} is extraordinarily useful as
it allows us to include the resummation of higher order QCD effects
that come from physically well--understood region of (almost)
parallel emission of partons off the quark or antiquark coming
from the primary $\gamma^*\rightarrow q\overline{q}$ splitting.
\end{itemize}
For the virtual photon, as opposed to the real one, its PDF
\footnote{More precisely their pointlike parts.}
can therefore be regarded as ``merely'' describing higher order
perturbative effects and not its ``genuine'' structure. Although this
distinction between the content of PDF of real and virtual photons
exists, it does not affect the extraordinary {\em phenomenological}
usefulness of PDF of the virtual photon. As shown in \cite{prd}
the nontrivial part of the contributions of resolved transverse
virtual photon ($\gamma_T^*$) to NLO calculations of dijet
production at HERA is large and affects significantly the conclusions
of phenomenological analyses of existing experimental data.

\section{PDF of $\gamma_L^*$ in QED}
\label{sec:qed}
Most of the present knowledge of the structure of the photon
comes from experiments at ep and e$^+$e$^-$ colliders, where
the incoming leptons act as sources of transverse and longitudinal
virtual photons of virtuality $P^2$ and momentum fraction $y$.
To order $\alpha$ their respective unintegrated fluxes are given as
\begin{eqnarray}
f^{\gamma^*_T}(y,P^2) & = & \frac{\alpha}{2\pi}
\left(\frac{1+(1-y)^2)}{y}\frac{1}{P^2}-\frac{2m_{\mathrm e}
^2 y}{P^4}\right),
\label{fluxT} \\
f^{\gamma^*_L}(y,P^2) & = & \frac{\alpha}{2\pi}
\frac{2(1-y)}{y}\frac{1}{P^2}.
\label{fluxL}
\end{eqnarray}
Phenomenological analyses of interactions of virtual photons and
their PDF have so far concentrated on its transverse polarization.
Neglecting longitudinal photons is a good approximation for
$y\rightarrow 1$, where $f^{\gamma^*_L}(y,P^2)\rightarrow 0$, as
well as for small virtualities $P^2$, where PDF of $\gamma_L^*$
vanish by gauge invariance. But how small is ``small'' in fact? For
instance, should we take into account the contribution of
$\gamma^*_{L}$ to jet cross--section in the region $E_T\gtrsim 5$
GeV, $P^2\gtrsim 1$ GeV$^2$, where most of the data on virtual
photons obtained in ep collisions at HERA come from? The present
paper is devoted to addressing this and related questions.

In pure QED and to order $\alpha$ the probability of finding inside
$\gamma_L^*$ of virtuality $P^2$ a quark with mass $m_q$, charge $e_q$,
momentum fraction $x$ and virtuality $\tau\le M^2$, is given, in units
of $3e_q^2\alpha/2\pi$, as \cite{prd}
\begin{equation}
q^{\mathrm {QED}}_L(x,m_q^2,P^2,M^2)=
\frac{4x^2(1-x)P^2}{\tau^{\mathrm {min}}}
\left(1-\frac{\tau^{\mathrm {min}}}{M^2}\right),
\label{fullresult}
\end{equation}
where $\tau^{\mathrm {min}}=xP^2+m_q^2/(1-x)$.
The quantity defined in (\ref{fullresult}) has a clear physical
interpretation: it describes the flux of quarks that are almost
collinear with the incoming photon and ``live'' longer
than $1/M$. For $\tau^{\mathrm {min}}\ll
M^2$ the expression (\ref{fullresult}) simplifies to
\begin{displaymath}
q^{\mathrm {QED}}_L(x,m_q^2,P^2,M^2)=
\frac{4x^2(1-x)P^2}{xP^2+m_q^2/(1-x)},
\end{displaymath}
which for $x(1-x)P^2\gg m_q^2$ further reduces to
\begin{equation}
q^{\mathrm {QED}}_L(x,0,P^2,M^2)=4x(1-x).
\label{virtualL}
\end{equation}
whereas for $x(1-x)P^2\ll m_q^2$
\begin{displaymath}
q^{\mathrm {QED}}_L(x,m_q^2,P^2,M^2)\rightarrow
\frac{P^2}{m_q^2}4x^2(1-x)^2
\end{displaymath}
demonstrating the fact that in QED the onset of $\gamma_L^*$ is governed
by the ratio $P^2/m_q^2$.

\section{QCD improved PDF of $\gamma_L^*$}
\label{sec:qcd}
QCD improved PDF of $\gamma_L^*$ have been derived in the
leading-logarithmic approximation and for $1\lesssim P^2\ll M^2$ in
\cite{plb}. By ``leading--log'' we mean resummation of the terms
$(\alpha_s\ln M^2)^k$ at each order $k$ of perturbative QCD. Note
that for $\gamma_T^*$ there is one power of $\ln M^2$ more at each
order of $\alpha_s$, the additional one coming from the primary
QED $\gamma^*\rightarrow q\overline{q}$ splitting. In the case of
$\gamma_L^*$ the analogous splitting gives rise to the term
(\ref{virtualL}), which is constant in $P^2$. The resulting
expressions
\footnote{The parameterization of PDF of $\gamma_L^*$ can be
obtained from chyla@fzu.cz.}
exhibit typical hadronic form of scale dependence and contain
$\Lambda_{\mathrm{QCD}}$ as the only free parameter. The condition
$P^2\ll M^2$ guarantees clear physical
meaning of the resulting quark and gluon distribution functions.
Moreover, by staying away from the region $P^2\sim M^2$ we avoid
the region where power corrections of the type $P^2/M^2$ are
essential and, in fact, more important than the effects described
by PDF. The restriction from below $1~{\mathrm{GeV}}^2\lesssim P^2$
ensures that hadronic parts of PDF of $\gamma_L^*$,
which have not been taken into account in the derivation in \cite{plb},
can be safely neglected with respect to the pointlike ones
\footnote{This claim is based on experience with SaS sets
of parameterizations \cite{sas1} and the assumption that hadronic
parts of PDF of $\gamma_L^*$ can be related to those of
$\gamma_T^*$ \cite{plb}.}.

The relevance of resolved $\gamma_L^*$ in hard collisions of virtual
photons
\footnote{$\gamma_L^*$ contributes to soft collisions and related
quantities, like $\sigma_{\mathrm{tot}}(\gamma^*{\mathrm{p}})$, as well,
but we restrict our discussion to hard collisions.
For the former, the reader is referred to \cite{friberg2}.}
depends on the theoretical framework one works in. In the next two
Sections we shall discuss the effects of including resolved
$\gamma_L^*$ within the LO as well as NLO QCD calculations. The
difference between the numerical relevance of resolved
$\gamma_L^*$ in these two cases arises from the fact that
parton level calculations contain at the order $\alpha^2\alpha_s^2$
some of the effects that go into the definitions of quark distribution
function of $\gamma_T^*$ and $\gamma_L^*$.

\section{Resolved $\gamma_L^*$ in LO QCD calculations}
\label{sec:LO}
\subsection{DIS on $\gamma^*$}
\label{subsec:dis}
\begin{figure}[b]\unitlength=1mm
\begin{picture}(110,70)
\put(0,35){\epsfig{file=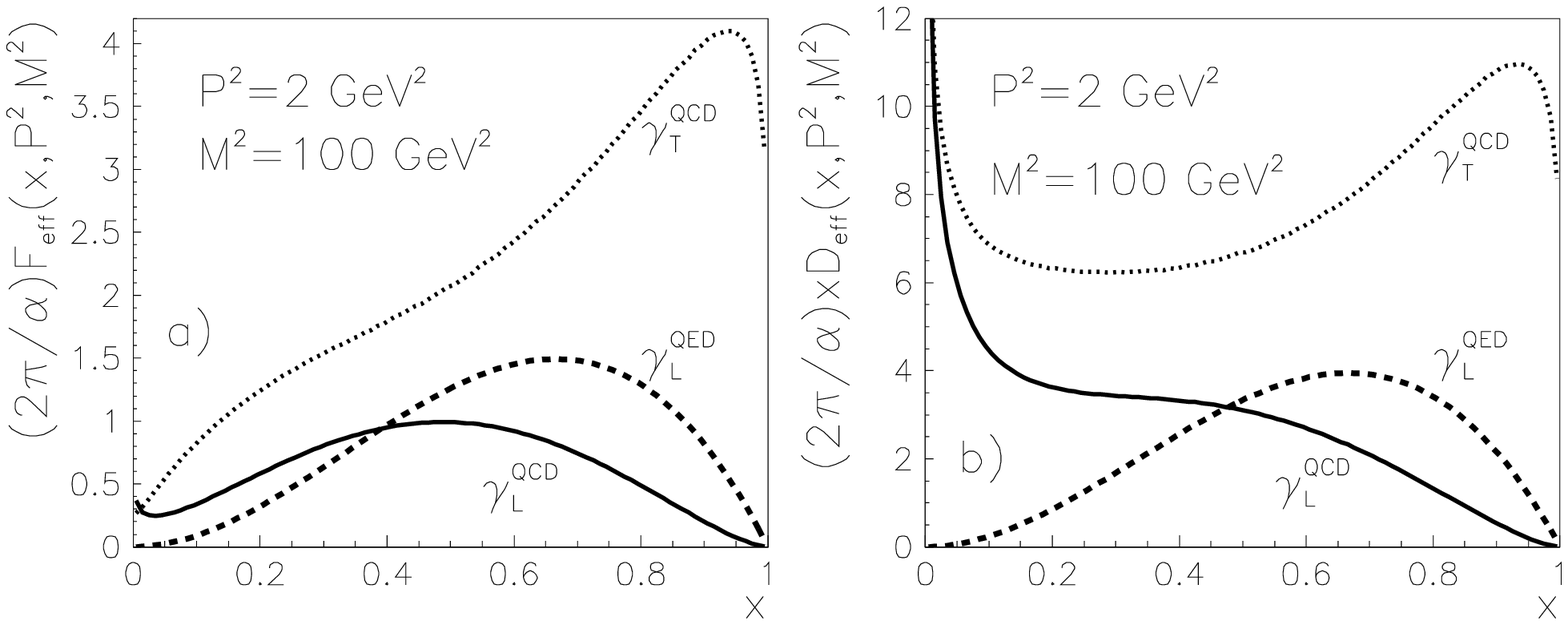,width=8.7cm}}
\put(0,0){\epsfig{file=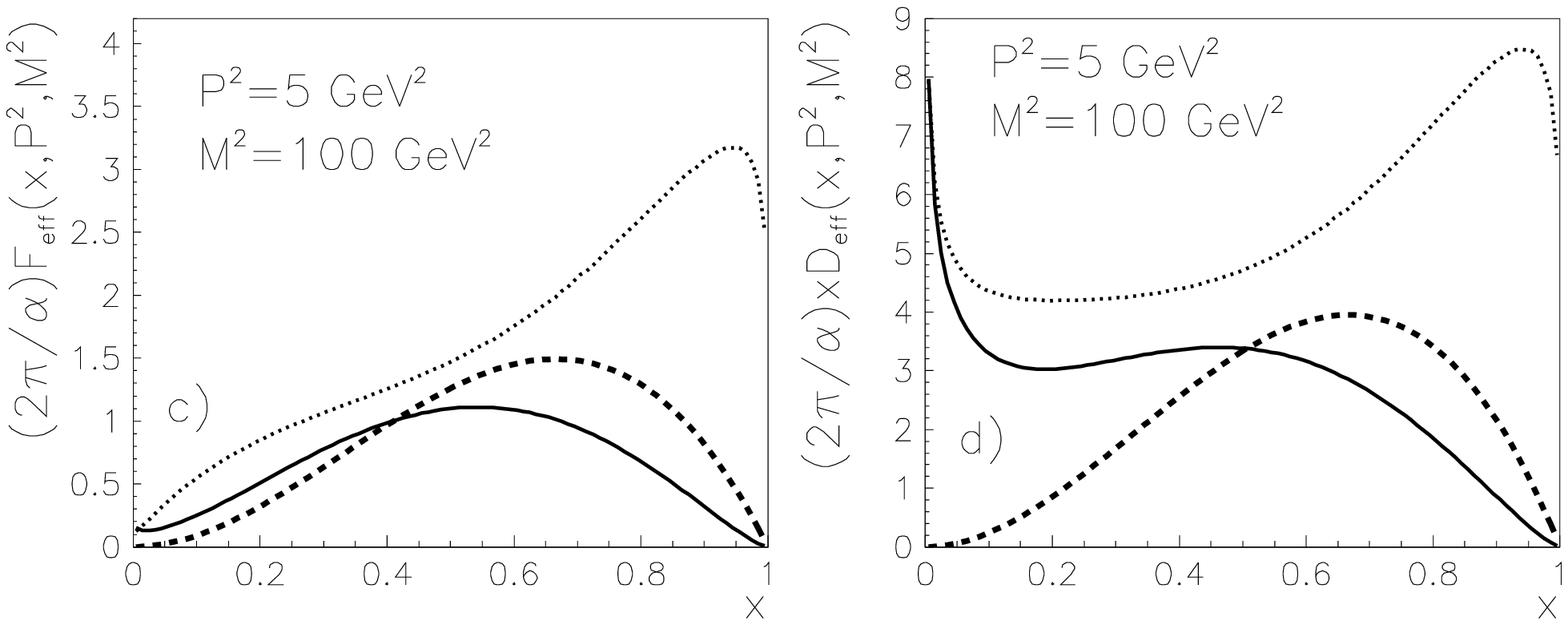,width=8.7cm}}
\end{picture}
\caption{Comparison of the contributions of $\gamma_T^*$ and
$\gamma_L^*$ to $F^{\gamma}_{\mathrm{eff}}$ (left) and
$D_{\mathrm{eff}}$ (right) for $P^2=2,5$ GeV$^2$ and $M^2=100$
GeV$^2$.}
\label{eff2}
\end{figure}
In LO QCD the structure function $F_2^{\gamma}$ of the virtual
photon is given in terms of quark distribution
functions by the same expression as for hadrons
\footnote{We disregard the consequences of the reformulation of
QCD analysis of $F_2^{\gamma}$ proposed by one of us in
\cite{jfactor} as they do not concern the main point of our
discussion.}
\begin{displaymath}
F_2^{\gamma}(x,P^2,Q^2)=  \sum_{i}2x e_i^2
\left(q_i(x,P^2,Q^2)+\overline{q}_i(x,P^2,Q^2)\right).
\end{displaymath}
In all existing phenomenological analyses only target $\gamma^*_T$
has been taken into account, despite the fact that for $P^2\ll Q^2$
experiments at LEP \cite{L3,OPAL} actually measure
\footnote{Neglecting the difference of the fluxes
(\ref{fluxT}-\ref{fluxL}), which is a good approximation at
small $y$, typical for LEP experiments.}
the ``effective'' structure function
\begin{displaymath}
F_{\mathrm{eff}}^{\gamma}(x,P^2,Q^2)\equiv
F_{2,T}^{\gamma}(x,P^2,Q^2)+F_{2,L}^{\gamma}(x,P^2,Q^2)
\end{displaymath}
given as the sum of contributions from target $\gamma_L^*$ and
$\gamma_L^*$. In Fig. \ref{eff2}a,c we compare, for two pairs of
$P^2$ and $Q^2$ typical for current experiments at LEP,
$F_2^{\gamma}$ obtained with SaS1D parameterization \cite{sas1} of
PDF of $\gamma_T^*$ with the contributions from target $\gamma_L^*$
evaluated using both the QED and QCD expressions for $q_L(x,P^2,M^2)$
discussed in the preceding two Sections. The contributions from
$q_L^{\mathrm{QED}}$ peak
around $x\simeq 0.7$, with QCD effects suppressing them at large $x$
and enhancing them on the other hand for $x\lesssim 0.4$. The presence
of the term proportional to $\ln M^2$ in the expression for $q_T$ in
both QED and QCD implies the dominance of $\gamma_T^*$ at large $M^2$,
but one would have to go to very large $M^2$ for $\gamma_L^*$ to become
negligible with respect to $\gamma_T^*$. For fixed $M^2$ the relative
importance of $\gamma_L^*$ with respect to $\gamma_T^*$ grows with
$P^2$, but to retain clear physical meaning of PDF we stay throughout
this paper in the region $P^2\ll M^2$.

\subsection{Dijet production in ep collisions}
\label{subsec:LOdijet}
The measurement of dijet production in ep collisions
provides another way of investigating interactions of virtual
photons \cite{H1eff,phd}. In general the corresponding cross sections
are given as sums of contributions of all possible parton level
subprocess. The simplest way of demonstrating the importance of
contributions of resolved $\gamma_L^*$ employs the approximation
\cite{Chris} in which dijet cross sections are expressed in terms of
a single {\em effective parton distribution function} of the photon
(either $\gamma_T^*$ or $\gamma_L^*$) defined as
\begin{eqnarray}
\lefteqn{D_{\mathrm{eff}}(x,P^2,M^2) \equiv} & & \nonumber\\
& & \sum_{i=1}^{n_f}\left(q_i(x,P^2,M^2)+
\overline{q}_i(x,P^2,M^2)\right)+\frac{9}{4}G(x,P^2,M^2), \nonumber
\label{deff}
\end{eqnarray}
\begin{figure*}[t]\sidecaption
\epsfig{file=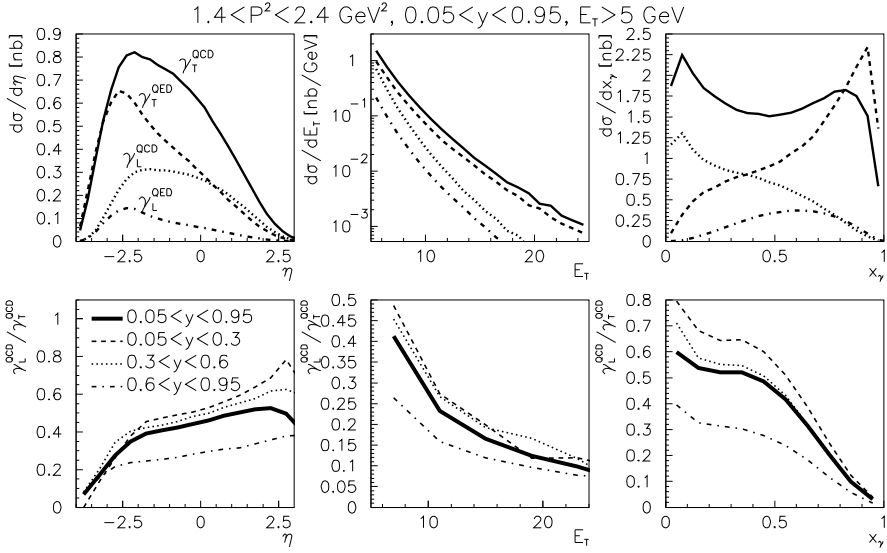,width=12cm}
\caption{Upper three plots:
diparton cross sections, corresponding to target $\gamma_T^*$ and
$\gamma_L^*$ and using QED as well as QCD improved PDF of the
latter, plotted as functions of $\eta,E_T$ and $x_{\gamma}$ for
$1.4\le P^2\le 2.4$ GeV$^2$, $0.05\le y\le 0.95$,
$E_T\ge 5$ GeV, without any restriction on $\eta$.
Lower three plots: ratia of the contributions of resolved
$\gamma^*_L$ (using PDF of \cite{plb}) to those of $\gamma_T^*$
(evaluated with PDF of \cite{sas1}), integrated over the whole region
$0.05\le y\le 0.95$, as well as in three indicated subintervals.}
\label{qcd11}
\end{figure*}
where the factorization scale $M$ is conventionally identified with
(a multiple of) jet $E_T$: $M=\kappa E_T$.
In Figs. \ref{eff2}b,d the contributions to $D_{\mathrm{eff}}$ from
$\gamma_T^*$ and $\gamma_L^*$ are compared for two pairs of
$P^2$ and $M^2$ typical for HERA experiments. In addition to effects at
large $x$, which are similar to those for $F_{\mathrm{eff}}^{\gamma}$,
$D_{\mathrm{eff}}$ gets a sizable contribution from $\gamma_L^*$
at small $x$, coming from its gluon content. The rise
of $D_{\mathrm{eff}}$ at small $x$ is particularly clear effect of
QCD improved PDF of $\gamma_L^*$.
\begin{figure*}\sidecaption
\epsfig{file=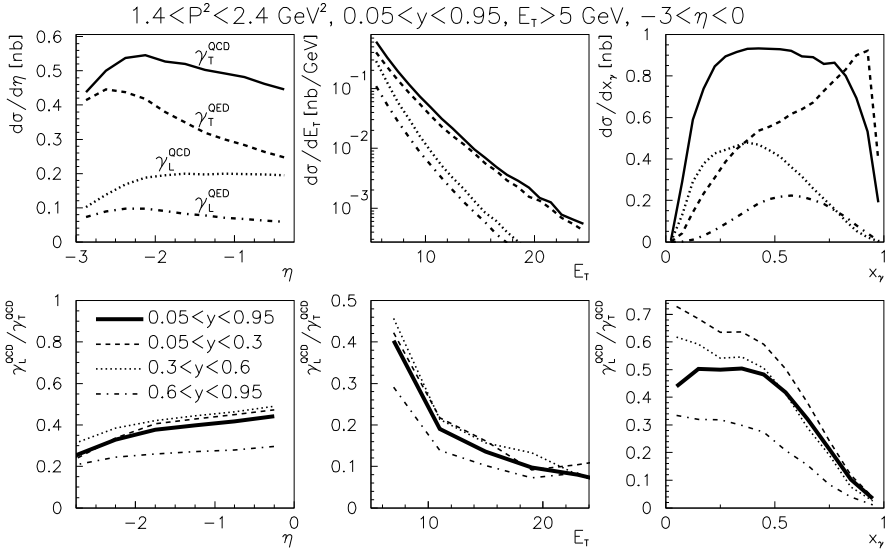,width=12cm}
\caption{The same as in Fig. \ref{qcd11}, but for experimentally
motivated restricted region $-3\le\eta\le 0$.}
\label{limqcd11}
\end{figure*}
After this estimate, we now proceed to discuss the
contributions of $\gamma_L^*$ to dijet cross sections evaluated with
HERWIG 5.9 event generator at the parton level. We could have
used for this purpose also JETVIP \cite{JETVIP}, which we shall use
later at the NLO, but using HERWIG at the LO allows us to
\begin{itemize}
\item estimate hadronization effects,
\item cross--check the modifications implemented in JETVIP in
order to include the effects of $\gamma_L^*$.
\end{itemize}
For the purpose of this study we have modified standard HERWIG 5.9 by
adding the option of generating the flux of $\gamma_L^*$ combined
with the call to QED or QCD improved PDF of $\gamma_L^*$. For
$\gamma_T^*$ the SaS1D PDF were used.
All calculations were performed for $0.05\le y\le 0.95$, three
windows of $P^2$: $1.4\le P^2\le 2.4~{\mathrm {GeV}}^2$,
$2.4\le P^2\le 4.4~{\mathrm {GeV}}^2$ and
$4.4\le P^2\le 10~{\mathrm{GeV}}^2$ and the following cuts on
parton $E_T$
$$E_T^{(1)},E_T^{(2)}\ge E_T^c,
~E_T^c=5,10~{\mathrm{GeV}}.$$
The effects of H1 and ZEUS detector acceptances have been approximately
taken into account by performing all calculations without any
restriction on parton pseudorapidity as well as for $-3\le \eta\le 0$.

The results for the first window in $P^2$ and without the cuts on
$\eta$ are presented as functions of $\eta,x_{\gamma}$ and
$E_T$ in Fig. \ref{qcd11}.
The characteristic dependence of the contributions of resolved
$\gamma_L^*$ on $y$ is illustrated by plotting for each of the
distributions in $\eta,E_T$ and $x_{\gamma}$ also its ratio to
that of $\gamma_T^*$ for the whole interval $0.05\le y\le 0.95$, as
well as for three indicated subintervals. Except for $x_{\gamma}$
close to $1$, QCD improved PDF of $\gamma_L^*$ enhance its
contributions to dijet cross sections compared to those based on
the purely QED. For $y\lesssim 0.5$ and $x_{\gamma}$ close to $0$
or $\eta\simeq 2.5$, the contributions of resolved $\gamma_L^*$
amount to about $80\%$ of those of $\gamma_T^*$, whereas on average
this number is around $50\%$. Reducing the range of $\eta$ to
$-3\le\eta\le 0$ affects (see Fig. \ref{limqcd11}) mainly the
distribution ${\mathrm{d}}\sigma/{\mathrm{d}}x_{\gamma}$ by
suppressing it at both endpoints $x_{\gamma}=0$ and $x_{\gamma}=1$.
The ratia of the contributions of $\gamma_L^*$ and $\gamma_T^*$ are,
however, affected only little by this cut.
\begin{figure*}[t]\sidecaption
\epsfig{file=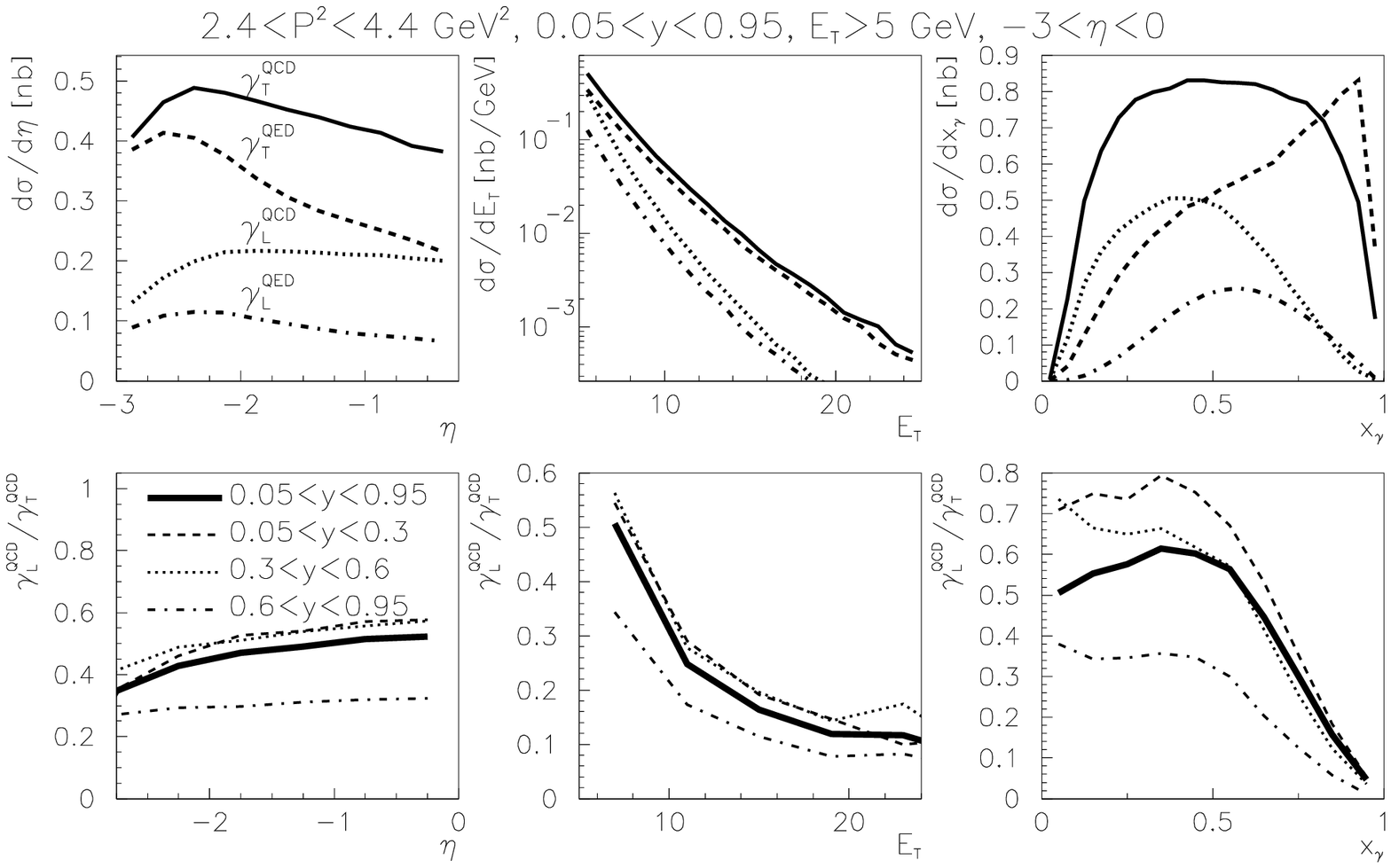,width=12cm}
\caption{The same as in Fig. \ref{limqcd11} but
for $2.4\le P^2\le 4.4$ GeV$^2$.}
\label{limqcd21}
\end{figure*}
\begin{figure*}\sidecaption
\epsfig{file=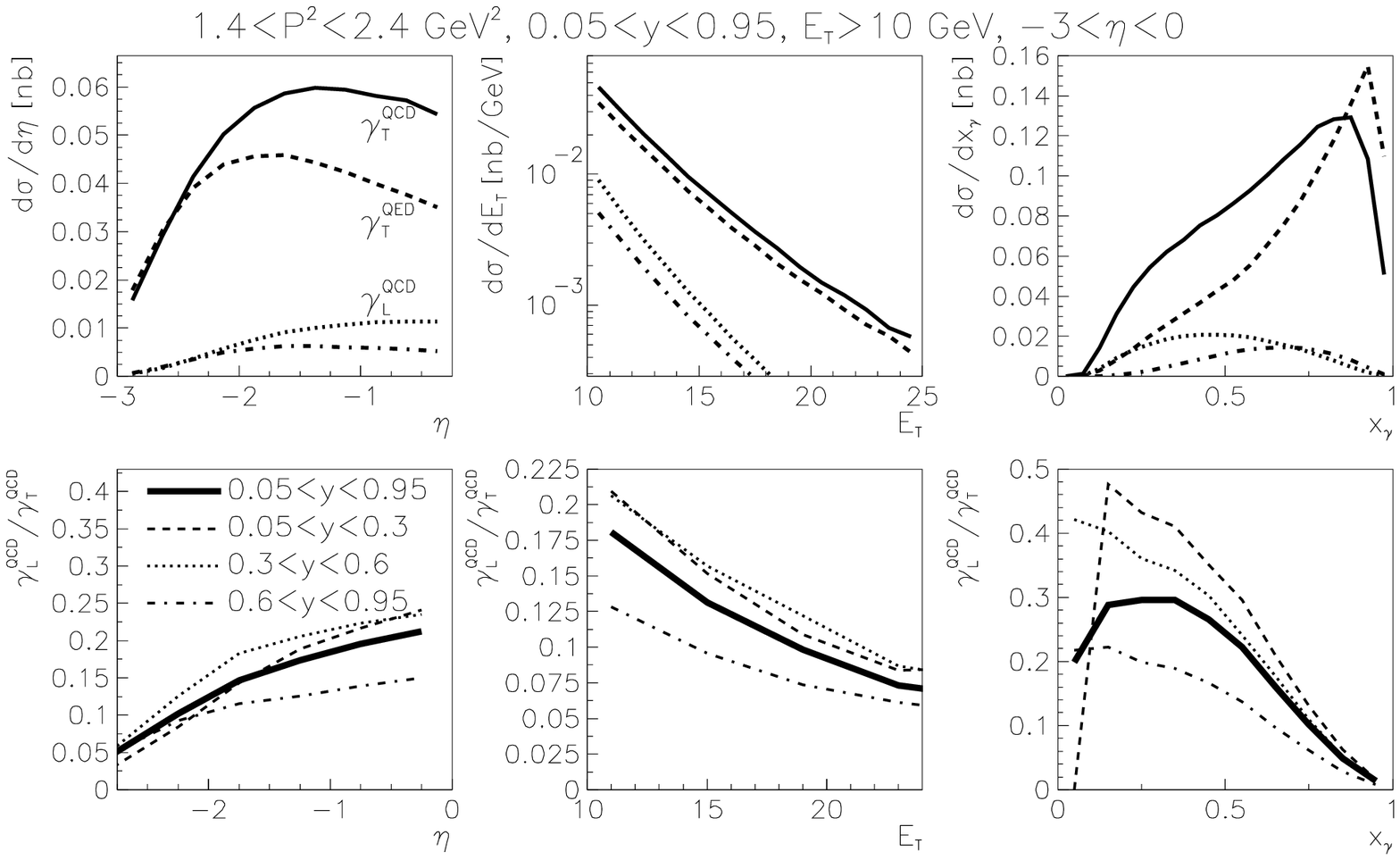,width=12cm}
\caption{The same as in Fig. \ref{limqcd11} but for $E_T^c=10$ GeV.}
\label{limqcd110}
\end{figure*}
Increasing the photon virtuality enhances, as shown in Fig.
\ref{limqcd21}, the relative importance of resolved $\gamma_L^*$
with respect to $\gamma_T^*$ On the contrary, rising the threshold
$E_T^c$ from $5$ GeV to $10$ GeV reduces it, as illustrated in Fig.
\ref{limqcd110}, by a factor of about 2, since large $E_T$ require
large $x_{\gamma}$, where quarks from $\gamma_T^*$ dominate.

Summarizing the message of Figs. \ref{qcd11}-\ref{limqcd21}, we
conclude that in the region $\Lambda^2\ll P^2\ll M^2\approx E_T^2$
the contributions of $\gamma_L^*$ are substantial, particularly for
\begin{itemize}
\item small $y$,
\item low $E_T$,
\item $x_{\gamma}\lesssim 0.5$, corresponding to
$\eta$ close to the upper edge.
\end{itemize}
The cuts enforced by H1 and ZEUS acceptances reduce the sensitivity
to $\gamma_L^*$, but its contributions still make up typically
$30-50\%$ of those of $\gamma_T^*$ and can be identified by
their characteristic $y$ and $P^2$ dependencies.

\section{Resolved $\gamma_L^*$ in NLO QCD calculations}
\label{sec:NLO}
\begin{figure}[t]
\resizebox{0.5\textwidth}{!}{\includegraphics{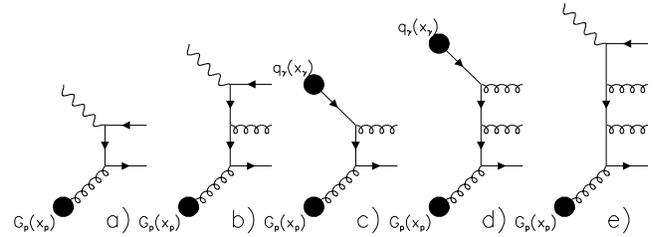}}
\caption{Examples of diagrams contributing to dijet production in
ep collisions at the orders $\alpha\alpha_s$ (a), $\alpha\alpha_s^2$
(b,c), and $\alpha\alpha_s^3$ (d,e) taking into account that the upper
blobs representing quark distribution functions of the photon are
proportional to $\alpha$.}
\label{diagrams}
\end{figure}

The relevance of resolved $\gamma_L^*$ within the framework of NLO
parton level calculations of dijet cross sections in ep collisions has
been investigated using JETVIP \cite{JETVIP}, the only NLO parton level
MC program including both direct and resolved photon contributions. In
specifying the powers of $\alpha$ and $\alpha_s$ corresponding to
various Feynman diagrams we discard one common power of $\alpha$ coming
from the vertex where the virtual photon is emitted by the incoming
electron. This vertex is also left out in diagrams of
Fig. \ref{diagrams}.

JETVIP contains full set of partonic cross sections for the direct
photon contributions up the order $\alpha\alpha_s^2$. Examples of the
corresponding diagrams are shown in Fig. \ref{diagrams}a,b. To go one
order of $\alpha_s$ higher and perform complete calculation of the direct
photon contributions up to order $\alpha\alpha_s^3$ would require
evaluating tree diagrams like that in Fig. \ref{diagrams}e, as well
as one--loop corrections to diagrams like in Fig. \ref{diagrams}b and
two--loop corrections to diagrams like in Fig. \ref{diagrams}a.
So far, such calculations are not available. In addition to complete
${\cal O}(\alpha\alpha_s^2)$ direct photon contributions JETVIP
includes also the resolved photon ones with partonic cross sections up
to the order $\alpha_s^3$, exemplified by diagrams in Fig.
\ref{diagrams}c,d.
The rationale for including in the resolved channel terms of the order
$\alpha_s^3$ is discussed in detail in \cite{prd}. Once the concept of
virtual photon structure is introduced, part of the direct photon
contribution (which for the virtual photon is actually nonsingular) is
subtracted and included in the definition of PDF of $\gamma^*$. To avoid
misunderstanding we shall
henceforth use the term ``direct unsubtracted'' (DIR$_{\mathrm{uns}}$)
to denote NLO direct photon contributions {\em before} this subtraction,
reserving the term ``direct'' (DIR) for the results {\em after} it. In
this terminology the complete JETVIP calculations are given by the sum
of direct and resolved parts and denoted DIR$+$RES. In JETVIP only the
convolution of QED splitting term (plus some finite terms) corresponding
to $\gamma_T^*$
\begin{equation}
q^{\mathrm{QED}}_T(x,P^2,M^2)=
\frac{\alpha}{2\pi}3e_q^2\left(x^2+(1-x)^2\right)\ln\frac{M^2}{xP^2}.
\label{splitterm}
\end{equation}
with $\alpha_s^2$ partonic cross sections are subtracted from
DIR$_{\mathrm{uns}}$ calculations.
We recall that in any NLO DIR$_{\mathrm{uns}}$ calculation
both $\gamma_T^*$ and $\gamma_L^*$ are taken into account exactly up to
the order $\alpha\alpha_s^2$. Introducing the concept of resolved
$\gamma_T^*$ and $\gamma_L^*$ implies the replacement of the
convolution (denoted $\sigma({\mathrm{PSP}})$) of photon splitting
terms ((\ref{splitterm}) for $\gamma_T^*$ and (\ref{virtualL}) for
$\gamma_L^*$) and order $\alpha_s^2$ partonic cross sections with the
contribution (denoted $\sigma_{T,L}({\mathrm{RES}})$) of the resolved
$\gamma^*_{T,L}$. The net effect of this operation is thus the
addition to $\sigma({\mathrm{DIR}}_{\mathrm{uns}})$ of the differences
$\Delta_{T,L}\equiv\sigma_{T,L}({\mathrm{RES}})-
\sigma_{T,L}({\mathrm{PSP}})$
\begin{equation}
\sigma({\mathrm{DIR}}+{\mathrm{RES}})=
\sigma({\mathrm{DIR}_{\mathrm{uns}}})+\Delta_T+\Delta_L.
\label{tot}
\end{equation}
The appropriate measure of the relevance of resolved
$\gamma_L^*$ in NLO calculations is thus the ratio
\begin{equation}
r_{LT}^{\mathrm{NLO}}(E_T,\eta)\equiv
\frac{\Delta_L(E_T,\eta)}{\Delta_T(E_T,\eta)}.
\label{rnontriv}
\end{equation}
Note that as for the LO QCD calculations the corresponding measure is
the ratio $\sigma_L({\mathrm{RES}})/\sigma_T({\mathrm{RES}})$, the
relevance of $\gamma_L^*$ in hard collisions is in general different at
LO and NLO orders. For $\gamma_L^*$ the cross section
$\sigma_L({\mathrm{RES}})$ is given by the convolution of QCD improved
PDF of $\gamma_L^*$ with partonic cross sections up to the order
$\alpha_s^3$.

To include the effects of resolved $\gamma_L^*$, we have modified
JETVIP with the help from Bj\"{o}rn P\"{o}tter in three places by
adding:
\begin{itemize}
\item the flux (\ref{fluxL}) of $\gamma_L^*$,
\item the photon splitting term (\ref{virtualL}) corresponding to
$\gamma_L^*$,
\item the call to PDF of $\gamma_L^*$.
\end{itemize}
We have checked our modifications against HERWIG as
well as internally within JETVIP. In the first case we compared LO
JETVIP results for $\gamma_L^*$ with analogous results obtained with
HERWIG 5.9 for the same QCD improved PDF of initial $\gamma_L^*$.
Taking into account small differences between the way JETVIP and
HERWIG
\begin{itemize}
\item set the scale of PDF and $\alpha_s$,
\item treat (light) quark mass effects,
\item reconstructs kinematics from generated $x_{\gamma}$,
\end{itemize}
we have found very satisfactory agreement in both shape and absolute
normalization of resulting distributions in all three variables
$x_{\gamma},\eta$ and $E_T$.

The modification of the photon splitting term (\ref{splitterm})
to the form appropriate for $\gamma_L^*$ has been checked by
comparing JETVIP results for $\sigma_L({\mathrm{PSP}})$ with LO JETVIP
results in the resolved channel obtained with purely QED expression
(\ref{virtualL}) for light quark distribution functions. Apart from
the opposite sign, the latter should be equal to the former as, indeed,
it turned out to be the case to within a few $\%$.

\subsection{Hadronization corrections}
\label{subsec:hadronization}
Any meaningful comparison of JETVIP results with experimental data must
involve estimates of the effects describing the conversion of partons
to hadrons. These hadronization corrections are not simple to define,
but adopting the definition used by experimentalists \cite{Wobisch},
we have found \cite{phd} that they depended sensitively and in
correlated manner on the pseudorapidity and transverse energy of jets.
For $E_T^c=5$ GeV, hadronization corrections become large and strongly
model dependent for $\eta\lesssim -2.5$. We have therefore
restricted our analysis to $-2.5\le\eta\le 0$, where they are flat in
$\eta$ and do not exceed $10$\%.

\subsection{Results}
\label{subsec:JETVIP}
\begin{figure}
\epsfig{file=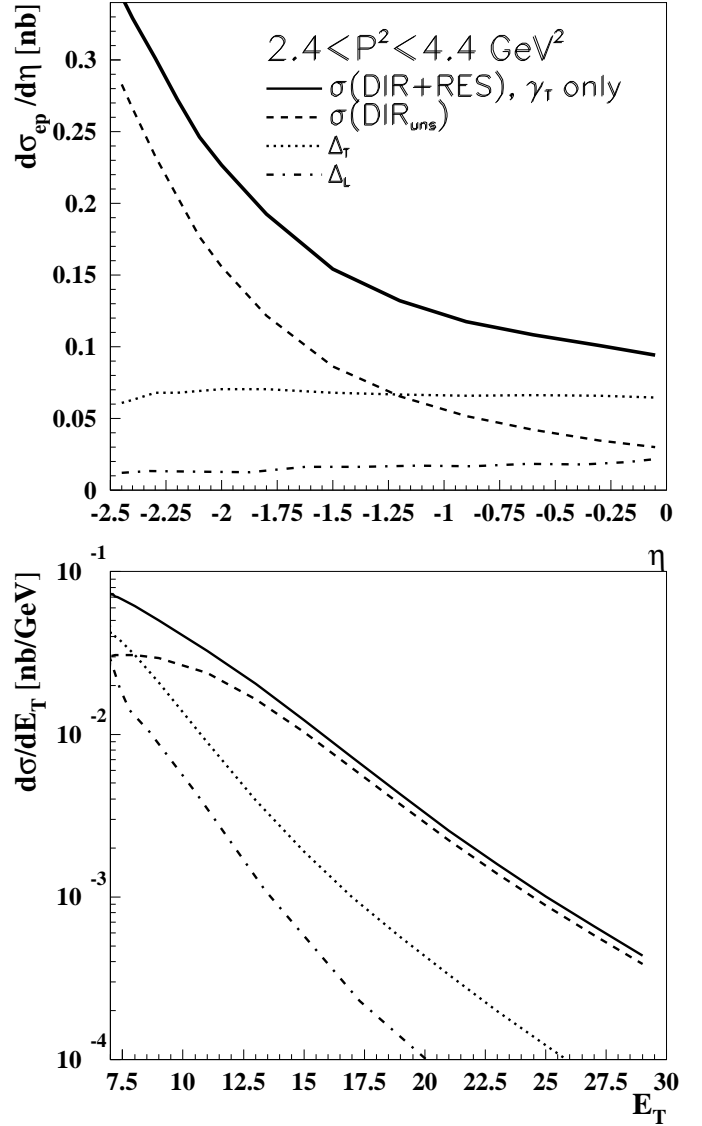,width=9cm}
\caption{Comparison of nontrivial parts $\Delta_T$ and $\Delta_L$
of the contributions of $\gamma_T^*$ and
$\gamma_L^*$ to  ${\mathrm{d}}\sigma/{\mathrm{d}}\eta$ and
${\mathrm{d}}\sigma/{\mathrm{d}}E_T$ distributions. The results
of direct unsubtracted and full calculations using in the resolved
channel $\gamma_T^*$ only are shown as well.}
\label{okno2}
\end{figure}
We have redone the calculation of \cite{long} using QCD improved
PDF of $\gamma_L^*$, but otherwise with the same assumptions
concerning renormalization and factorization scales
\footnote{In PDF of $\gamma_L^*$ we set $\Lambda^2_{\mathrm{QCD}}=0.1$
GeV$^2$.} and for
identical kinematical region $$-2.5\le \eta\le 0,~~E_T^{(1)}\ge
7,~E_T^{(2)}\ge 5~\mathrm{GeV}.$$ The resulting distributions
${\mathrm{d}}\sigma/{\mathrm{d}}\eta$ and
${\mathrm{d}}\sigma/{\mathrm{d}}E_T$ corresponding to the second
window in $P^2$ are shown in Fig. \ref{okno2}. We plot there
separately all three contributions on the r.h.s. of (\ref{tot}),
as well as their sum defined in (\ref{tot}) but including the
contributions of $\gamma_T^*$ only. Note that both $\Delta_T$ and
$\Delta_L$ are almost flat in $\eta$ and rapidly falling in $E_T$,
the latter fall-off being faster for $\Delta_L$ as expected due to
harder shape of PDF of $\gamma_T^*$. The resulting
$r_{LT}^{\mathrm{NLO}}(E_T,\eta)$ rises slowly from about 0.2 at
$\eta=-2.5$ to 0.35 at $\eta=0$, but decreases appreciably with
$E_T$. Integrated over $E_T$, we find
$r_{LT}^{\mathrm{NLO}}(\eta)\simeq 0.3$, but for $E_T$ close to
the lower cut--off $E_T^c=7$ GeV, this ratio increases to about
0.5. Note also that for $\eta$ close to $\eta\simeq 0$,
$\Delta_L$ approaches the results of DIR$_{\mathrm{uns}}$
calculations.

\begin{figure}
\epsfig{file=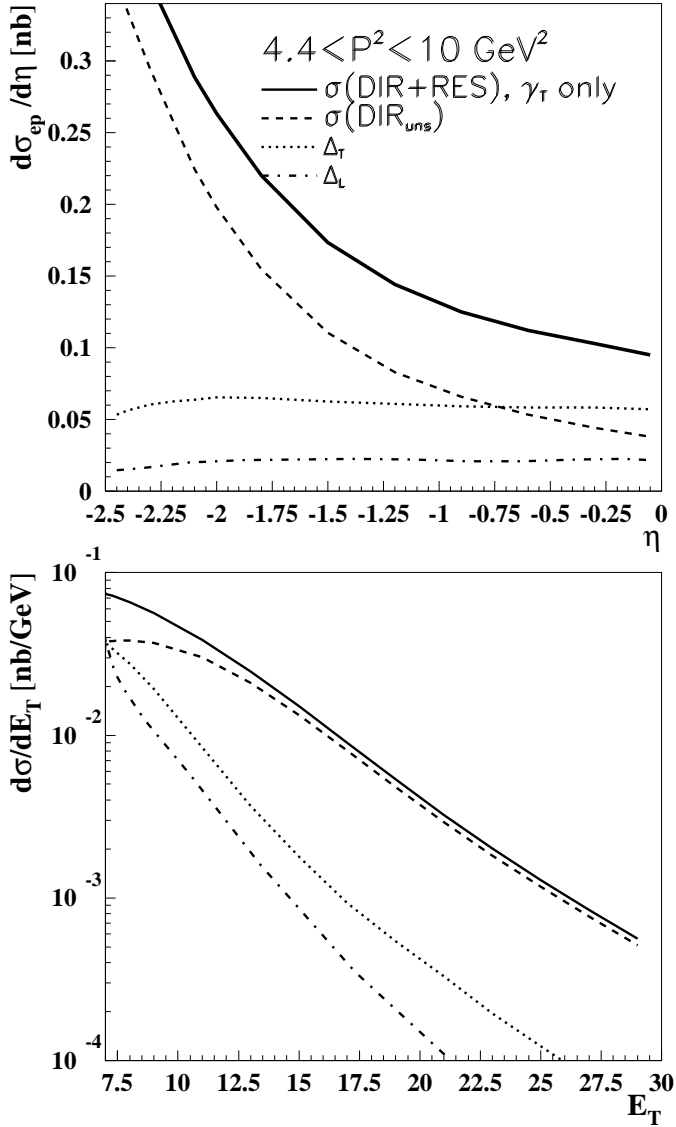,width=9cm}
\caption{The same as in Fig. \ref{okno2} but for
$4.4\le P^2\le 10$ GeV$^2$.}
\label{okno3}
\end{figure}

Increasing the photon virtuality:
\begin{itemize}
\item reduces the relevance of resolved $\gamma_T^*$ as measured
by the ratio $\Delta_T/\sigma({\mathrm{DIR}}_{\mathrm{uns}})$, but
\item increases the relative importance of resolved
$\gamma_L^*$ with respect to resolved $\gamma_T^*$ as measured by
the ratio $r^{\mathrm{NLO}}_{LT}$.
\end{itemize}
This is illustrated in Fig. \ref{okno3}, which shows the same plots
as in Fig. \ref{okno2} but for $4.4\le P^2\le 10$ GeV$^2$. In this
interval the mean value of $r_{LT}^{\mathrm{NLO}}$ is about $0.38$,
but for $E_T$ close to $E_T^c=5$ GeV it approaches unity.
Rising the cut--off $E_T^c$ reduces the
relevance of $\gamma_L^*$ with respect to $\gamma_T^*$, for much
the same reasons as in LO calculations.

In general, the relative importance of resolved $\gamma_T^*$ and
$\gamma_L^*$ is determined by two circumstances: the presence of
``large log'' $\ln(M^2/P^2)$ in PDF of $\gamma_T^*$ and the
difference in shapes of PDF of $\gamma_T^*$ and $\gamma_L^*$.
At very large value of the ratio $M^2/P^2$ the first effect
is clearly more important and leads to dominance of resolved
$\gamma_T^*$. However, in presently accessible range at HERA this
``large log'' is fairly small number around 3 and
thus the fact that PDF of $\gamma_T^*$ are harder than those of
$\gamma_L^*$ plays equally important role.

Inclusion of the contributions of resolved $\gamma_L^*$ in
phenomenological analyses of HERA data on dijet production
helps bring the theoretical predictions closer to the H1 data
\cite{phd}, but a thorough analysis of the evidence for
resolved $\gamma_L^*$ in current HERA data requires detailed
discussion of a number of points, and is beyond the scope of
this paper.

\section{Summary and conclusions}
We have analyzed the contributions of resolved $\gamma_L^*$ to
virtual photon structure function $F_{\mathrm{eff}}^{\gamma}$ and
dijet cross sections measured at HERA, using
the recently constructed parameterization of QCD improved PDF of
$\gamma_L^*$. The contributions of resolved $\gamma_L^*$ were shown
to be nonnegligible with respect to those of $\gamma_T^*$, but their
relevance depends on the order of QCD calculations employed and
kinematical region considered. Within the LO QCD and in the
kinematical regions accessible at LEP and HERA, they amount
typically to $40-50$\% of those coming from resolved $\gamma_T^*$,
but in parts of phase space (small $y$ and $x_{\gamma}$ or low
$E_T$) this number is even larger. Within the NLO calculations
of virtual photon interactions the relative importance of
$\gamma_L^*$ with respect to $\gamma_T^*$ is smaller, but still
clearly of phenomenological relevance. In both cases the effects
of QCD improved PDF of $\gamma_L^*$ are clearly observable.

\begin{acknowledgement}
We are grateful to J. Cvach, C. Friberg and B. P\"{o}tter for
interesting discussions concerning the structure and interactions of
longitudinal virtual photons and to B. P\"{o}tter for help in modifying
JETVIP. This work was supported in part by Grant Agency of the Academy of
Sciences of the Czech Republic under the grants No. A1010821 and
B1010005.
\end{acknowledgement}

\end{document}